\documentclass[10pt,aps,prb,twocolumn,showpacs,amssymb,floatfix]{revtex4-1}
\usepackage{amsmath,graphics,epsfig,mathrsfs,physics,bbold}
\usepackage{amsfonts}
\usepackage{amssymb}
\usepackage{graphicx}
\usepackage{bm}
\newcommand*{\Tensor}[1]{\overline{\overline{#1}}}
\newcommand{\etal}{{\em et al.~}}

\usepackage{color}
\usepackage{hyperref}
\hypersetup{backref=true,pagebackref=true,hyperindex=true,colorlinks=true,citecolor=blue, breaklinks=true,urlcolor=
blue,linkcolor=blue,bookmarks=true,bookmarksopen=true}

\begin{document}
\title{
Quantum Anomalous Hall Effect and Anderson Chern Insulating Regime in Noncollinear Antiferromagnetic 3Q State
}
\author{Papa Birame Ndiaye$^{1}$}
\email{papabirame.ndiaye@kaust.edu.sa}
\author{Adel Abbout$^{1}$}
\author{Aur\'{e}lien Manchon$^{1,3}$}
\email{aurelien.manchon@kaust.edu.sa}
\affiliation{$^1$King Abdullah University of Science and Technology (KAUST), Physical Science and Engineering Division (PSE), Thuwal 23955-6900, Saudi Arabia.}
\affiliation{$^3$King Abdullah University of Science and Technology (KAUST), Computer, Electrical and Mathematical Science and Engineering Division (CEMSE), Thuwal 23955-6900, Saudi Arabia.}
\date{\today}

\begin{abstract}  
We investigate the emergence of both quantum anomalous Hall and disorder-induced Anderson Chern insulating phases in two dimensional hexagonal lattices, with antiferromagnetically ordered 3Q state and in the absence of spin-orbit coupling. Using tight-binding modeling, we show that such systems display not only a spin-polarized edge-localized current, the chirality of which is energy dependent but also an impurity-induced transition from trivial metallic to topological insulating regimes, through one edge mode plateau. We compute the gaps' phase diagrams, and demonstrate the robustness of the edge channel against deformation and disorder. Our study hints at the 3Q state as a promising building block for dissipationless spintronics based on antiferromagnets. 
\end{abstract}

\pacs{72.25.-b,73.43.-f,75.50.Ee}
\maketitle
\section{Introduction} Since the control of the antiferromagnetic order parameter by electrical means has been demonstrated \cite{Zelezny2014,Wadley2016}, antiferromagnets have undoubtedly emerged as credible candidates for the replacement of ferromagnets as the active and upgrading spin-dependent element on which spintronic devices are based. With their numerous outstanding properties, they provide a rich playground to study unique magnetic properties combined with unconventional transport phenomena \cite{Jungwirth2016,Baltz2018}.
In particular, the interplay between electronic transport, topological properties of the ground states (in reciprocal space) and antiferromagnetic order in real space opens auspicious perspectives in the field of topological antiferromagnetic spintronics \cite{Smejkal2018}. Indeed, although antiferromagnets break time-reversal symmetry locally, they are invariant under the combination of spin rotation and crystal symmetry operation (e.g., lattice translation in G-type collinear antiferromagnets, mirror symmetry in kagome lattice etc.), which provides an analog to Kramers' degeneracy theorem. A direct consequence is that antiferromagnetism preserves the gapless states of topological materials such as topological insulators \cite{Mong2010,Ghosh2017} and Weyl semimetals \cite{Tang2016,Smejkal2017}. \par

The search for topological effects has been particularly fruitful in the context of non-collinear antiferromagnets. For instance, anomalous Hall effect in non-collinear but {\em coplanar} antiferromagnets has been recently observed \cite{Nakatsuji2015,Nayak2016}, and attributed to the synergistic coaction of the Berry phase of the electronic ground states in the presence of spin-orbit coupling (SOC) and symmetry breaking due to the non-collinear spin texture \cite{Chen2014,Kubler2014}. Along similar lines of thought, large longitudinal and transverse spin currents in non-collinear coplanar antiferromagnets have also been reported \cite{Zelezny2017}. 
Because of they do not necessitate SOC, non-collinear, {\em non-coplanar} antiferromagnets constitute an appealing platform to sustain topological and trivial states and realize phase transitions between them. As a matter of fact, non-coplanar magnetic moments distributed on a lattice promote the emergence of spin-Berry phase even in the absence of SOC \cite{Taguchi2001} and can experience topological phase transitions \cite{Ohgushi2000,Shindou2001}.

Topological phase transitions in condensed matter have been intensively scrutinized over decades \cite{Kosterlitz2017,Haldane2017}. Such transitions are accompanied by the emergence of localized edge states unaffected by disorder and immune to backscattering, resulting in quantum (spin or anomalous) Hall effects in topological insulators \cite{Hasan2010}. Quantum anomalous Hall (QAH) effect is characterized by frictionless edge states in the absence of magnetic field. One way to fulfill this quantum state is to start with a time-reversal symmetric ($\mathbb{Z}_2$) topological insulator displaying quantum spin Hall effect, and break time reversal symmetry by introducing magnetic order by magnetic doping \cite{Chang2013,Zhang2014} or surface hybridization with a magnetic insulator \cite{Luo2013,Katmis2016}. Alternatively, Haldane showed that such a state can be obtained by engineering the band structure in such a way that conduction electrons experience a local, staggered magnetic flux that vanishes globally \cite{Haldane1988,Qiao2014}. Such a situation can be achieved with the Kondo-lattice model on a triangular lattice \cite{Martin2008,Venderbos2012} or with the double-exchange model on a kagome lattice \cite{ Ishizuka2013,Chern2014} where non-collinear, non-coplanar antiferromagnetism is stabilized and provides both non-vanishing Berry curvature and orbital gap.

\begin{figure}[t]
\includegraphics[width=8.4cm]{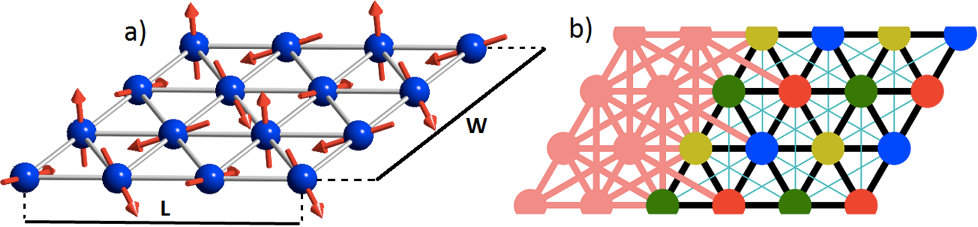}
\caption{\label{fig1} (Color online) (a) 3Q spin structure in a 2D triangular lattice. 
(b) \{Lead + System\}: in green, red, blue, yellow the four magnetic atoms of the motif. The bold black (green) lines are the first (second) nearest neighbor couplings.}
\end{figure}

In the present work, we exploit the non-collinear, noncoplanar antiferromagnetic texture in the absence of spin-orbit coupling to realize topologically protected edge transport. Taking the two-dimensional (2D) triangular lattice with 3Q spin texture as prototype model, we show that spin-polarized QAH effect is achieved without SOC in zero net magnetization material. The chirality of the topological edge modes depends on the gap in which the Fermi energy is located. They are shown to be robust against disorder and unaffected by geometrical defects inside the lattice. We also investigate the rich next-nearest-neighbor hopping phase diagram for the topological band gaps, opening around the high symmetry Dirac points. Finally, we demonstrate that gradually tuned disorder can drive the 3Q antiferromagnetic metallic state into a topological Anderson Chern insulating phase \cite{Li2009,Groth2009}, characterized by a conductance plateau $e^2/h$.

\section{Triangular lattice model} Let us start with a 2D hexagonal lattice underpinning the peculiar 3Q antiferromagnetic structure as depicted in Fig. \ref{fig1} (a): its chiral 3D spin configuration, equivalent to a superposition of three spin spirals 
is noncollinear, noncoplanar and nontrivial in real space, while exhibiting no overall net magnetization. Such a magnetic state was originally predicted by Kurz et al. \cite{Kurz2001} to emerge at the Mn/Cu(111) interface even in the absence of SOC, based on {\em ab-initio} calculations. The authors attributed the onset of the 3Q state the so-called 4-spin interaction, i.e., beyond the first nearest neighbor approximation. Further investigations in the context of the Kondo lattice and Hubbard model have indeed confirmed the thermodynamic stability of such a 3Q spin texture in the triangular lattice \cite{Martin2008,Kato2010,Venderbos2012}, pointing out the importance of itinerant electron contributions. The unit cell of the 3Q state has four chemically identical sites as seen in Fig. \ref{fig1} (a) with $\sum_{i=1}^4 {\bf m}_i={\bf 0}$. Nevertheless, the spin texture of the 3Q state gives rise to a nonvanishing spin chirality $\kappa={\bf m}_i\cdot({\bf m}_j\times {\bf m}_k) \ne 0$ \cite{Ohgushi2000}, even though the sum of these chiralities over the magnetic cell is zero, like in Haldane model \cite{Haldane1988}. The spin-Berry curvature $\kappa$ and its associated effective magnetic flux \cite{Shindou2001} induce anomalous Hall effect in frustrated ferromagnets \cite{Taguchi2001}, skyrmionic materials \cite{Yanagihara2002} but also 3Q antiferromagnets with no overall magnetism \cite{Martin2008}.

\begin{figure}[t]
\includegraphics[width=8.9cm]{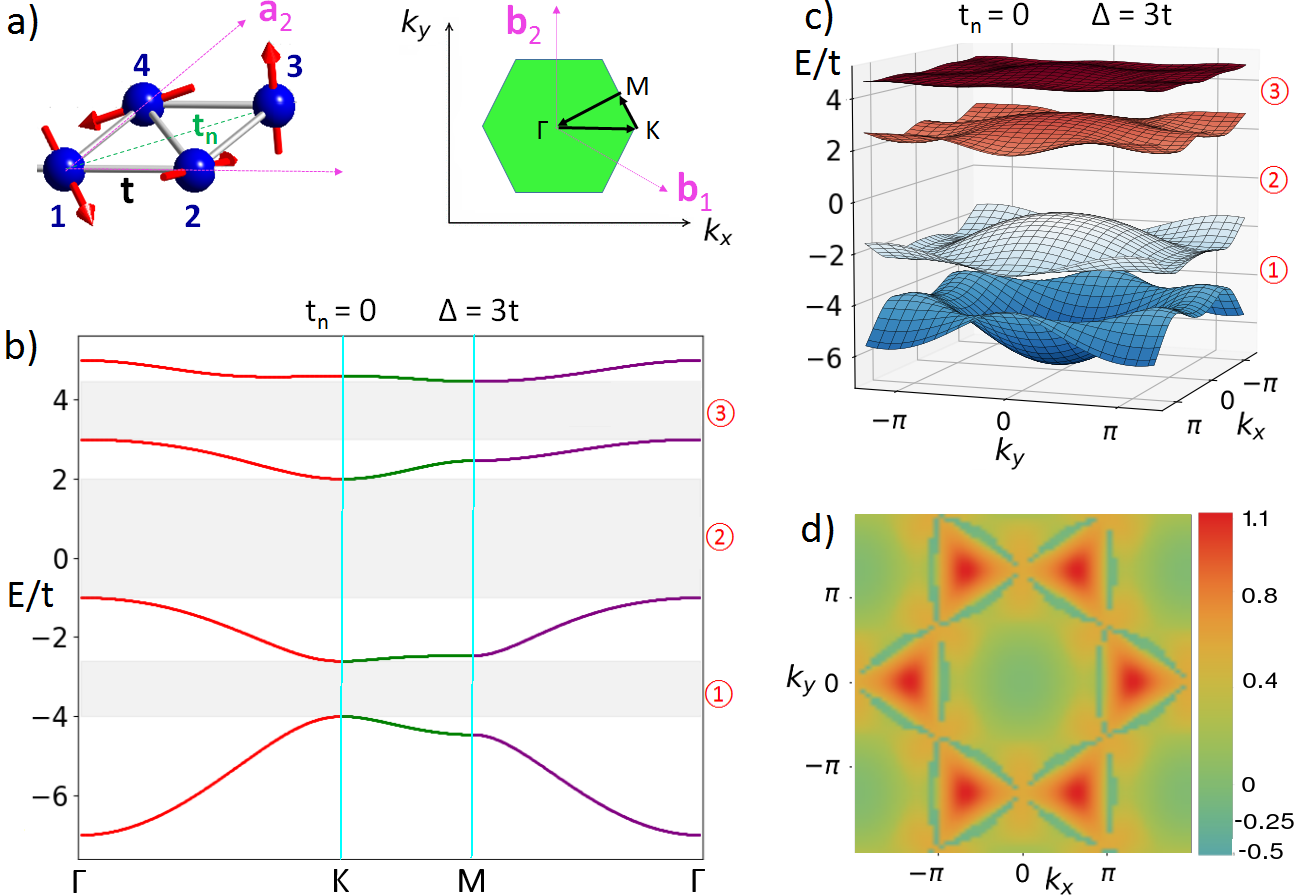}
\caption{\label{fig2} (Color online) (a) The unit cell with $\sum_{i=1}^4{\bf m}_i={\bf 0}$ and the first Brillouin zone. (b), (c) Band structure of the model in Eq. \eqref{Hk} with $t_n=0$, $\Delta=3t$. (d) Trace of the Berry curvature tensor $\text{Tr}[\Tensor{\Omega}_z({\bf k})]$ of the lowest degenerate band in (c) plotted on [$-2\pi$, $2\pi$]$\times$[$-2\pi$, $2\pi$].}
\end{figure}

The hexagonal lattice with the primitive cell formed by ${\bf a}_1=(1,0)$ and ${\bf a}_2=(1/2,\sqrt{3}/2)$ and the first Brillouin zone with the reciprocal lattice spanned by ${\bf b}_1=(2\pi,-2\pi/\sqrt{3})$ and ${\bf b}_1=(0,4\pi/\sqrt{3})$ are both depicted in Fig. \ref{fig2} (a). The single electron Hamiltonian for a tight-binding s-d model is 
\begin{multline}\label{Hr}
  \hat{H}=\sum_{i_{\alpha\beta\gamma\delta}}\hat{c}_{i_{\alpha\beta\gamma\delta}}^\dagger\big(\epsilon_{i_{\alpha\beta\gamma\delta}}+\Delta_{i_{\alpha\beta\gamma\delta}}
  {\bf m}_{i_{\alpha\beta\gamma\delta}}\cdot \hat{\boldsymbol{\sigma}} \big)\hat{c}_{j_{\alpha\beta\gamma\delta}} \\
  - t\sum_{\langle i_\alpha,j_\beta\rangle_{\gamma\delta}}\hat{c}_{i_{\alpha\gamma\delta}}^\dagger\hat{c}_{j_{\beta\gamma\delta}} 
  - t_n\sum_{\langle\langle i_\alpha,j_\gamma\rangle\rangle_{\beta\delta}} \hat{c}_{i_{\alpha\beta\delta}}^\dagger\hat{c}_{j_{\gamma\beta\delta}} 
  \end{multline}
where $\sum_{\langle i,j\rangle}$ and $\sum_{\langle\langle i,j\rangle\rangle}$ are the sum over the nearest neighbors and the next-nearest neighbor pairs with the hopping parameters $t$ and $t_n$, respectively. $\hat{\boldsymbol{\sigma}}$ is the vector of Pauli matrices and the indices $\alpha$, $\beta$, $\delta$ and $\gamma$ represent the positions of the atoms in the motif. The eigenvalue problem can be recast in the momentum space as ${\hat H}_{\bf k}\ket{u_{n{\bf k}}} = \varepsilon_{n{\bf k}}\ket{u_{n{\bf k}}}$, where ${\hat H}_{\bf k}= e^{-i{\bf k}\cdot {\bf r}}\hat{H}e^{i{\bf k}\cdot {\bf r}}$ is the $\bf k$-dependent Hamiltonian, $\varepsilon_{n{\bf k}}$ is the eigenenergy of the $n$th band and finally $\ket{u_{n{\bf k}}}$ is the periodic part of the Bloch wave function.
The Hamiltonian in momentum space straightforwardly reads
\begin{equation} \label{Hk}
\hat{H}_{\bf k}=\begin{bmatrix}
\Delta\hat{\boldsymbol{\sigma}}\cdot{\bf m}_1  &  K         & L_1         & L_2 \\
K           & \Delta\hat{\boldsymbol{\sigma}}\cdot{\bf m}_2 & L_2         & L_1 \\
L_1           & L_2          & \Delta\hat{\boldsymbol{\sigma}}\cdot{\bf m}_3& K \\
L_2           & L_1          & K          & \Delta\hat{\boldsymbol{\sigma}}\cdot{\bf m}_4
\end{bmatrix}
\end{equation}
with $K=-2t\cos\frac{1}{2}{\bf k}\cdot{\bf a}_1 - 2 t_n\cos\frac{1}{2}{\bf k}\cdot({\bf a}_1-2{\bf a}_2)$, $L_1=-2t\cos\frac{1}{2}{\bf k}\cdot({\bf a}_1-{\bf a}_2)- 2 t_n\cos\frac{1}{2}{\bf k}\cdot({\bf a}_1+{\bf a}_2)$, $L_2=-2t\cos\frac{1}{2}{\bf k}\cdot{\bf a}_2 - 2 t_n\cos\frac{1}{2}{\bf k}\cdot(2{\bf a}_1-{\bf a}_2)$, up to a multiplicative identity matrix $\mathbb{1}_2$.

\section{Spin-Berry phase-induced anomalous transport}

\subsection{Topological edge states} The band structure along the high symmetry points contour and its three dimensional version over the whole Brillouin zone are displayed in Fig. \ref{fig2} (b) and (c), respectively. There are four doubly degenerate bands, as dictated by Kramers' degeneracy theorem. Indeed, the four magnetic moments are oriented towards the four corners of a tetrahedron and performing a rotation ${\cal R}$ within this tetrahedron is equivalent to performing a translation $T_a$ in the crystal lattice. Therefore, the operation ${\cal O}={\cal R}T_a$ is a symmetry of the magnetic system and the operator ${\cal T}{\cal O}$ is antiunitary, $({\cal T}{\cal O})^2=-1$, implying double degeneracy of the bands. Besides the antiferromagnetic gap appearing in the middle (noted "gap 3"), two additional gaps are obtained in Figs. \ref{fig2} (b) and (c), referred to as "gap 1" for the lower one and "gap 3" for the upper one. 

\begin{figure}[b]
\includegraphics[width=8.8cm]{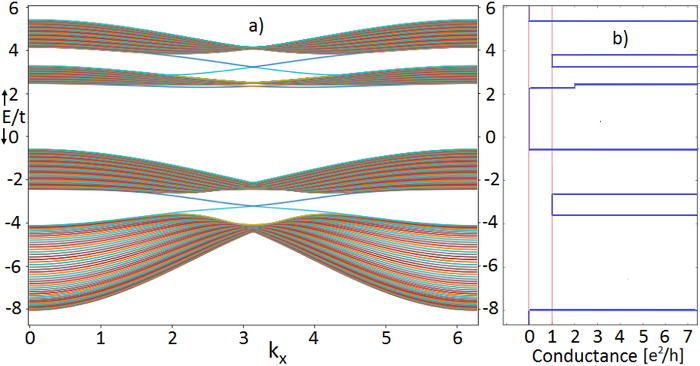}
\caption{\label{fig3} (Color online) 
(a) The band structure along the contour $\Gamma \rightarrow K \rightarrow \Gamma$ with the emergence of the gapless edge states seen in gaps 1 and 3, with $t_n=0.2t$, $\Delta=3t$. (b) Zoom on the conductance for the same parameters.}
\end{figure}

As mentioned in the introduction, the origin of the QAH effect is rooted in the existence of a non-vanishing k-space Berry curvature defined as ${\bm\Omega}({\bf k})=i\langle\nabla_{{\bf k}} u({\bf k}) |\times|\nabla_{{\bf k}} u({\bf k})\rangle$  for a non degenerate state. In case of degeneracy, a generalization to a matrix form is needed \cite{Shindou2005,Gradhand2012}. In fact, the time evolution of the system entails the occurrence of adiabatic transitions between states of the same subspace and therefore the Berry curvature tensor is constructed by all the wave function projections between the elements of the same subspace,

\begin{eqnarray} \Tensor{\bm\Omega}_{ij}({\bf k})&=&i\langle\nabla_{{\bf k}} u_i({\bf k}) |\times|\nabla_{{\bf k}} u_j({\bf k})\rangle\\
&&+i\sum_{l\in \zeta} \langle u_i({\bf k}) |\nabla_{{\bf k}} u_l({\bf k})\rangle \times \langle u_l({\bf k}) | \nabla_{{\bf k}} u_j({\bf k})\rangle,\nonumber
\end{eqnarray}

where $\zeta$ is the degenerate subspace. The second  term encodes the signature of the non Abelian topology of the covariant tensor. In our non-collinear, non-coplanar antiferromagnet, the bands are two fold degenerate and hence, the Berry curvature tensor for each degenerate band is a $2\times2$ non vanishing matrix. As a matter of fact, close to gaps $1$ and $3$, a non-vanishing k-space Berry curvature tensor emerges around $K$ points [see Fig. 2 (d)], enabling the onset of gapless chiral states at the edges. The physical origin of the Berry curvature is the exchange interaction that makes the spins of the conduction electrons align with the local moments, inducing a magnetic flux-like Berry curvature which gives rise to an anomalous velocity \cite{Taguchi2001,Shindou2001}. The Berry curvature tensor is not gauge invariant and this leads, when the gauge is not suitably chosen, to the appearance of jumps in the eigenvalues profile of the Berry curvature tensor \cite{Gradhand2012}. These jumps do not prevent from defining observables which are gauge invariant such the determinant $ \text{det}[\Tensor{\Omega}({\bf k})]$ and the trace $ \text{Tr}[\Tensor{\Omega}({\bf k})]$. For instance, in the insulating regime the anomalous Hall conductivity $\sigma_{xy}$ equals the integral of $ \text{Tr}[\Tensor{\Omega}({\bf k})]$ on the first Brillouin zone, i.e., 
\begin{eqnarray}
\sigma_{xy}&&=-(e^2/\hbar)\sum_n\int_{\rm BZ}d^2{\bf k}/(2\pi)^2 \text{Tr}[\Tensor{\Omega}_n({\bf k})]f_n({\bf k}),\\
&&=-(e^2/h){\cal C},\end{eqnarray} 
where $f_n({\bf k})$ is the Fermi distribution for the degenerate states, $n$ is the index of the degenerate subspace and ${\cal C}$ is the Chern number. The non-vanishing quantized $\sigma_{xy}$ in the gaps $1$ and $3$ is the signature of the non-trivial topology of the 3Q state which is exposed by the presence of edge states. This property is absent in collinear as well as in coplanar antiferromagnets.
\begin{figure}[t]
\includegraphics[width=8.6cm]{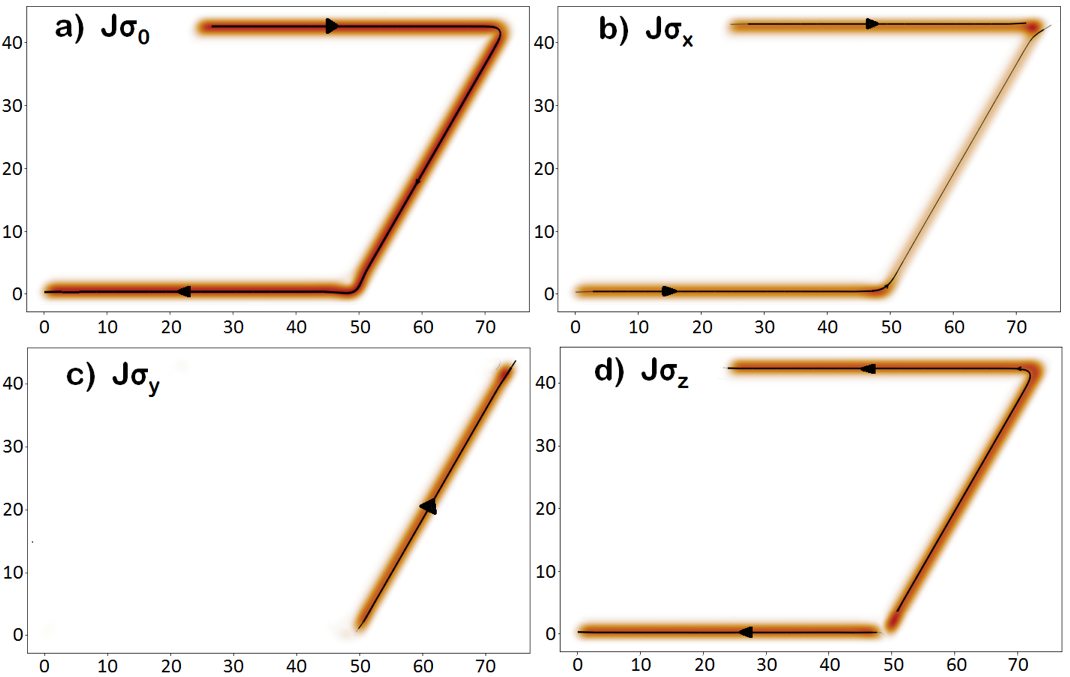}
\caption{\label{fig4} (Color online) 
(a)-(d) The unidirectional chiral edge current (for an energy lying in gap 1, with L = W = 50 and $t_n=0.2t$, $\Delta=3t$) and its spin-polarization. The black arrows show the direction of the current.}
\end{figure}

For further insight on the nature of these edge currents, we use the Kwant package \cite{Groth2014} to build a nanowire along the x-direction, as displayed in Fig. \ref{fig1}(b) and plot its one-dimensional band structure on Fig. \ref{fig3}(a) with the same parameters as in Fig. \ref{fig2}. The edge states clearly appear in the bottom and top gaps (gap 1 and gap 3, respectively), while the middle gap, gap 2, remains insulating. Fig. \ref{fig3}(b) displays the zoom of conductance when varying the conduction electron energy across the band structure. In the bulk bands, a large number of conducting channels are naturally seen because of the high value of the bulk density of states. In the gaps however, the situation is different: while gap 2 corresponds to zero conductance, gaps 1 and 3 correspond to one quantum of conductance, confirming that only one edge state conducts in these two cases.\par

To better characterize the nature of these edge states, we disconnect the right lead and compute the spatial profile of the charge current for electrons coming from the left lead and bouncing back to it. Fig. \ref{fig4}(a) shows this charge current when the operating energy is in the lower gap, gap 1. A chiral unidirectional edge state is propagating clockwise from the top edge to the bottom one, demonstrating the realization of the QAH effect in the absence of SOC. The same QAH effect is observed in the upper gap, gap 3, for which the single propagating edge mode has opposite chirality (anti-clockwise from the bottom edge to the top one). The associated spin currents ${\bf J}{\boldsymbol{\sigma}_{\rm x,y,z}}$ are plotted in Fig. \ref{fig4} (b), (c), (d): the charge current spin-polarization follows the average net magnetization of the edge line. For instance, on the top edge, the current is spin-polarized along ${\bf m}_1+{\bf m}_2$, while on the right edge it is spin-polarized along ${\bf m}_2+{\bf m}_3$ . 

\begin{figure}[b]
\includegraphics[width=8.4cm]{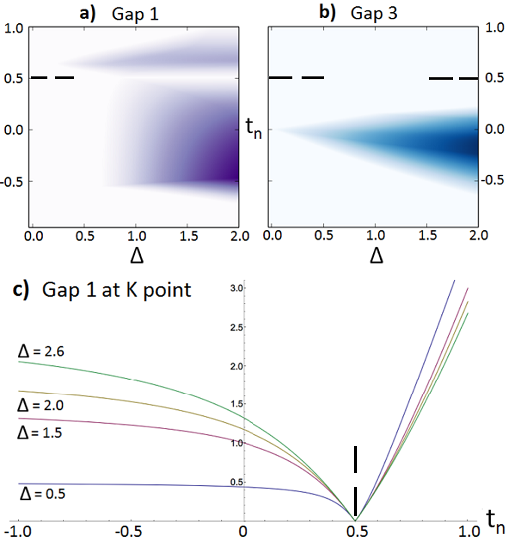}
\caption{\label{fig5} (Color online) 
(a), (b) Phase diagrams for the global gaps 1 and 3. $\Delta$ and $t_n$ are in units of $t$. (c) The local gap 1 at $K=(\pm 4\pi/3, 0)$, as a function of $t_n$, and for four $\Delta$. The Dirac point appears at $K$, for $t_n=0.5t$.}
\end{figure}

\subsection{Engineering the band structure} 

While each nearest neighbor triangular plaquette bears a spin Berry phase $\kappa={\bf m}_1\cdot({\bf m}_2\times{\bf m}_3)$, turning on the next-nearest neighbor hopping $t_n$ creates new triangular plaquettes with opposite flux $\kappa'={\bf m}_1\cdot({\bf m}_2\times{\bf m}_4)=-\kappa$ [see Fig. \ref{fig2}(a)]. Therefore, by tuning the relative strength between the nearest neighbor hopping $t$, the second nearest neighboring hopping $t_n$ and the exchange coupling $\Delta$, one can modify the emergent magnetic flux threading the magnetic structure and thereby engineer the band structure and its properties. For a fixed nearest neighbor hopping $t$, the conditions on the exchange coupling $\Delta$ and on the next-nearest neighbor $t_n$ (both in units of $t$) in order to get a topologically non-trival system are summarized in Fig. \ref{fig5} (a) and (b): the regions where a topological gap (gaps 1 and 3) exists are in colored scale, the darker the color is the larger the gap is. The gaps plotted in the phase diagram are global ones, they represent the difference between the lowest energy of the $n^{th}$ band and the highest energy of the $(n-1)^{th}$ band as computed over all the Brillouin zone. \par

The size of gap 1 at $K=(\pm4\pi/3, 0)$ is displayed in Fig. \ref{fig5} (c) for different exchange couplings $\Delta$ as a function of $t_n$: remarkably, gap 1 closes at $t_n=0.5t$, which corresponds to the appearance of a Dirac cone at $K$. This can be also viewed in Fig. \ref{fig5} (a), at the dashed line. When $t_n=0.5t$, the global gap 3 vanishes [see dashed line in Fig. \ref{fig5} (b)]. However, at this condition there is another Dirac cone at $K$, and a local and direct band crossing around the same energy range as gap 3. As a matter of fact, by expanding the energy dispersion close to $K$-point and around $t_n\approx 0.5t$, we obtain for these two different Dirac cones, $E_K^{\pm}(q)=\pm\Delta\pm\frac{\sqrt{3}}{4}ta|q|$, and $|q|$ is the norm of the momentum $q$ around the $K$ point.
Related to Fig. \ref{fig5}, notice that Fig. \ref{fig5} (c) shows a finite local gap when $\Delta=0.5t$ and for $-t<t_n<0.5t$ whereas Fig. \ref{fig5} (a) denies the existence of a global gap for these conditions. The bands are in fact overlapping and crossing indirectly at different $\bf k$-points of the Brillouin zone, therefore ensuring the existence of a local gap despite the absence of global gaps.\par

\begin{figure}[t]
\includegraphics[width=8.75cm]{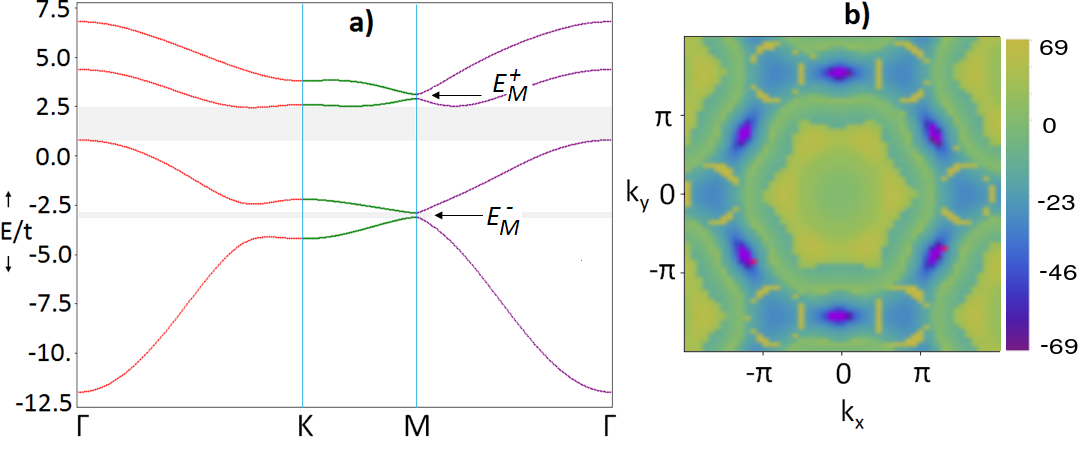}
\caption{\label{fig6} (Color online) 
(a) Band structure with $t_n=0.9t$ and $\Delta=3t$. (b) Trace of the Berry curvature tensor $\text{Tr}[\Tensor{\Omega}({\bf k})]$ of the lowest band for the same parameters, with the Dirac point $M$.}
\end{figure}

Another remarkable feature appears in gap 1 when $t_n\rightarrow t$. This time, the Dirac cones emerge at $M=(\pi, \pi/\sqrt{3})$, as seen in the band structure in Fig. \ref{fig6} (a) and located at the energy  $E_M^-$. The Berry curvature calculation of the lowest energy band, depicted in Fig. \ref{fig6} (b), confirms $M$ as the Dirac point around which the finite curvature is concentrated. The global gap 3 remains nonexistent for $t_n\rightarrow t$ as predicted in the phase diagram [Fig. \ref{fig5} (b)], but there is a Dirac-like crossing point at the energy $E_M^+$, as indicated in Fig. \ref{fig6} (a).

\begin{figure}[b]
\includegraphics[width=8.6cm]{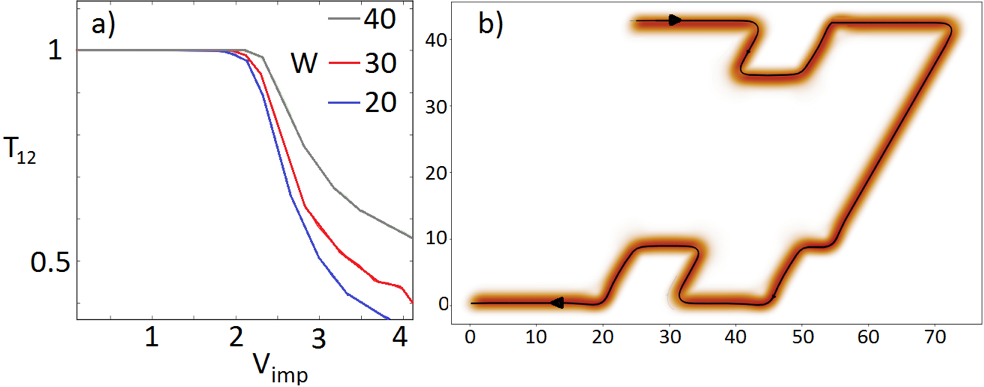}
\caption{\label{fig7} (Color online) 
(a) The transmission coefficient (energy in gap 1) vs disorder strength for three different widths, for L = 50. (b) Robustness of the unidirectional chiral edge current against the local geometrical defects.}
\end{figure}

\subsection{Robustness and Anderson Chern insulator} We now turn our attention towards the impact of Anderson disorder on the topological transport. For this purpose, we add to Eq. \eqref{Hr} nonmagnetic random on-site energy $V_0$, uniformly distributed on the range [$-V_{\rm imp}$, $V_{\rm imp}$] \cite{Ndiaye2017a}. The calculation of the conductance of the disordered strip is averaged over 1280 disorder configurations.
The normalized conductance as a function of disorder strength $V_{\rm imp}$ is plotted in Fig. \ref{fig7}(a) for three different widths $W$, and setting the value of Fermi energy in the topological gap 1 ($t_n=0.2t$, $\Delta=3t$).
The quantized conductance is robust for impurity strength ranging from 0 to $\sim2t$, showing that the topological edge currents are preserved even under relatively strong disorder. Above $V_{\rm imp}\approx 2t$, the conducting edge channel is progressively destroyed and the conductance starts decreasing significantly. In Fig. \ref{fig7}(b), we furthermore demonstrate that these QAH states persist under geometrical deformations of the system edges. The unidirectional transport enabled by the band topology is immune to the backscattering due to local geometrical defects or singularities. 

\begin{figure}[b]
\includegraphics[width=8.6cm]{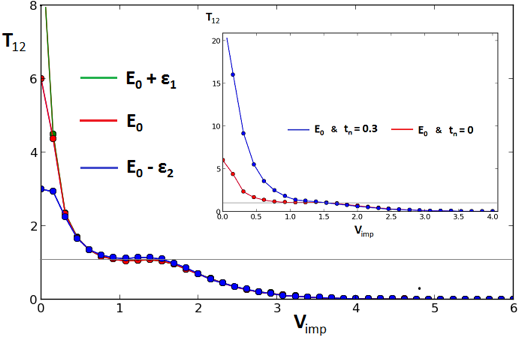}
\caption{\label{fig8} (Color online) 
(a) The transmission coefficient as a function of the disorder strength $V_{\rm imp}$, at fixed chemical potentials lying inside the bulk states above the gap 3 ($E_0=2.65t$, $\varepsilon_1=0.07t$ and $\varepsilon_1=0.04t$), the system parameters are W = L = 20 and $t_n=0$, $\Delta=t$. Inset: The comparison between conductance for two next-nearest neighbor hopping parameters 0 and 0.3$t$.}
\end{figure}

Finally, we show that in this antiferromagnetic 3Q state system in a metallic regime, disorder is capable of inducing the onset of another nontrivial quantum phase, termed topological Anderson Chern insulator. Anderson's theory of localization states that strong disorder can promote metal-insulator transition \cite{Anderson1958}. In systems possessing non-vanishing Berry curvature, the renormalization of the topological effective mass and chemical potential induced by Anderson disorder favors the emergence of topological edge states when transiting towards the insulating regime \cite{Groth2009,Li2009}. A similar effect is expected in the present case, where the non-trivial band structure topology is provided by the spin-Berry curvature.\par

Here, we consider the Fermi energy located within the bulk states, so that in the clean regime it is simply a topologically trivial metal with a finite number of bulk propagating modes. The behavior of the normalized conductance between the two leads as a function of disorder strength is plotted in Fig. \ref{fig8}, main panel, for three different Fermi energies fixed inside the bulk band structure ($E_0$, $E_0\pm\varepsilon_{1,2}$), and above gap 3. For the three cases, the conductance, high in the absence of disorder, decreases progressively upon increasing the disorder strength. Then, the three conductance curves merge and reach a plateau at $e^2/h$, featuring the emergence of one chiral transport channel flowing on the system's edge (bottom edge here because of the chirality in gap 3). This edge state is perfectly conducting in spite of the strong disorder, and its quantized plateau is credited to its topological robustness. Beyond a certain disorder strength, the three conductance curves drop to zero. To complete this study, the inset of Fig. \ref{fig8} compares the evolution of the conductance for different next-nearest neighbor hopping parameters, demonstrating that $t_n$ clearly influences the range of the disorder-induced QAH plateau by tuning the band structure topology as discussed above. We emphasize that while such a plateau has been originally observed in the context of $\mathbb{Z}_2$ topological insulators \cite{Groth2009,Li2009}, it was associated with quantum spin Hall effect. In the present case, the plateau is associated with QAH effect, i.e., charge edge currents.

\section{Conclusion} While non-collinear coplanar antiferromagnets have been recently proposed to host unconventional spin and anomalous currents \cite{Shindou2001,Chen2014,Kubler2014,Zelezny2017}, the present study highlights the potential of non-coplanar antiferromagnets for topological transport. 
In the former, the coexistence of SOC and non-collinear coplanar antiferromagnetic texture promote the both time-reversal symmetry breaking and non-vanishing Berry curvature. In contrast, in the latter the non-coplanar antiferromagnetic texture provides both ingredients, even in the absence of SOC \cite{Martin2008, Ishizuka2013,Chern2014,Venderbos2012}. The emergence of quantum phase of matter in frustrated quantum magnets has been the object of intense scrutiny in the past decades, with a particular focus on their magnetic phases and elementary excitations \cite{Batista2016}. For the experimental realization of QAH effect in antiferromagnets, magnetic pyrochlores \cite{Gardner2010} displaying all-in all-out spin configuration such as Pr$_2$Ir$_2$O$_7$ \cite{Machida2007} or Nd$_2$Ir$_2$O$_7$ \cite{Ma2015}, or layered triangular magnets such as cobaltites \cite{Ivanova2009}, and in particular Na$_{0.5}$CoO$_2$ \cite{Ning2008}, are considered as valuable candidates.

\acknowledgments
The authors thank Z. T. Ndiaye, G. E. W. Bauer, K.-J. Lee, X. R. Wang and F. Pi\'echon for valuable discussions. This work has been supported by the King Abdullah University of Science and Technology (KAUST) through the Office of Sponsored Research (OSR) [Grant Number OSR-2015-CRG4-2626].


\begin{thebibliography}{70}
\bibitem{Zelezny2014} J. Zelezny, H. Gao, K. Vyborny, J. Zemen, J. Masek, A. Manchon, J. Wunderlich, J. Sinova, and T. Jungwirth, Phys. Rev. Lett. {\bf113}, 157201 (2014).
\bibitem{Wadley2016} P. Wadley, \etal, Science {\bf351}, 587 (2016).
\bibitem{Jungwirth2016} T. Jungwirth, X. Marti, P. Wadley and J. Wunderlich, Nature Nanotechnology {\bf 11}, 231 (2016).
\bibitem{Baltz2018} V. Baltz, A. Manchon, M. Tsoi, T. Moriyama, T. Ono, Y. Tserkovnyak, Rev. Mod. Phys. {\bf90}, 015005 (2018).
\bibitem{Smejkal2018} L. Smejkal, Y. Mokrousov, Binghai Yan and A. H. MacDonald, Nature Physics {\bf14}, 242 (2018).
\bibitem{Mong2010} R. S. K. Mong, A. M. Essin, and J. E. Moore, Phys. Rev. B {\bf81}, 245209 (2010).
\bibitem{Ghosh2017} S. Ghosh and A. Manchon, Phys. Rev. B {\bf95}, 035422 (2017).
\bibitem{Tang2016} P. Tang, Q. Zhou, G. Xu and S.-C. Zhang, Nature Physics {\bf12}, 1100 (2016).
\bibitem{Smejkal2017} L. Smejkal, T. Jungwirth, J. Sinova, Phys. Status Solidi RRL {\bf11}, 1700044 (2017).
\bibitem{Nakatsuji2015} S. Nakatsuji, N. Kiyohara, T. Higo, Nature {\bf527}, 212-215 (2015).
\bibitem{Nayak2016} A. K. Nayak, J. E. Fischer, Y. Sun, B. Yan, J. Karel, A. C. Komarek, C. Shekhar, N. Kumar, W. Schnelle, J. K\"ubler, C. Felser, S. S. P. Parkin, Sci. Adv. 2016;2:e1501870 (2016).
\bibitem{Chen2014} H. Chen, Q. Niu, and A. H. MacDonald, Phys. Rev. Lett. {\bf112}, 017205 (2014).
\bibitem{Kubler2014} J. K\"ubler, C. Felser, Europhys. Lett. {\bf108}, 67001 (2014).
\bibitem{Zelezny2017} J. \v Zelezn\' y, Y. Zhang, C. Felser, and B. Yan, Phys. Rev. Lett. {\bf119}, 187204 (2017).
\bibitem{Taguchi2001} Y. Taguchi, Y. Oohara, H. Yoshizawa, N. Nagaosa, and Y. Tokura, Science {\bf29}, 2573 (2001).
\bibitem{Ohgushi2000} K. Ohgushi, S. Murakami, and N. Nagaosa, Phys. Rev. B {\bf62}, R6065 (2000).
\bibitem{Shindou2001} R. Shindou and N. Nagaosa, Phys. Rev. Lett. {\bf87}, 116801 (2001).
\bibitem{Kosterlitz2017} J. Kosterlitz, Rev. Mod. Phys. {\bf 89}, 040501 (2017).
\bibitem{Haldane2017} D. Haldane, Rev. Mod. Phys. {\bf 89}, 040502 (2017).
\bibitem{Hasan2010} D. Hsieh, D. Qian, L. Wray, Y. Xia, Y. Hor, R. J. Cava, and M. Z. Hasan, Nature (London) {\bf 452}, 970 (2008); M. Z. Hasan and C. L. Kane, Rev. Mod. Phys. {\bf 82}, 3045 (2010).
\bibitem{Chang2013} C.-Z. Chang, J. Zhang, X. Feng, J. Shen, Z. Zhang, M. Guo, K. Li, Y. Ou, P. Wei, L.-L. Wang, \etal Science {\bf340}, 167-170 (2013).
\bibitem{Zhang2014} H. Zhang, Y. Xu, J. Wang, K. Chang, and S.-C. Zhang, Phys. Rev. Lett. {\bf 112}, 216803 (2014).
\bibitem{Luo2013} W. Luo, and X.-L. Qi, Phys. Rev. B {\bf87}, 085431 (2013); S. V. Eremeev, V. N. Menshov, V. V. Tugushev, P. M. Echenique, and E. V. Chulkov, Phys. Rev. B {\bf88}, 144430 (2013).
\bibitem{Katmis2016} F. Katmis, V. Lauter, F. S. Nogueira, B. A. Assaf, M. E. Jamer, P. Wei, B. Satpati, J. W. Freeland, I. Eremin, D. Heiman, P. Jarillo-Herrero, and J. S. Moodera, Nature {\bf533}, 513 (2016).
\bibitem{Haldane1988} F. D. M. Haldane, Phys. Rev. Lett. {\bf61}, 2015 (1988).
\bibitem{Qiao2014} Z. Qiao, W. Ren, H. Chen, L. Bellaiche, Z. Zhang, A. H. MacDonald, and Q. Niu, Phys. Rev. Lett. {\bf112}, 116404 (2014).
\bibitem{Venderbos2012} J. W. F. Venderbos, M. Daghofer, J. van den Brink, and S. Kumar, Phys. Rev. Lett. {\bf109}, 166405 (2012).
\bibitem{Martin2008} I. Martin and C. D. Batista, Phys. Rev. Lett. {\bf101}, 156402 (2008).
\bibitem{Ishizuka2013} H. Ishizuka and Y. Motome, Phys. Rev. B {\bf87}, 081105(R) (2013).
\bibitem{Chern2014} G.-W. Chern, A. Rahmani, I. Martin, and C. D. Batista, Phys. Rev. B {\bf90}, 241102(R) (2014).
\bibitem{Li2009} J. Li, R. L. Chu, J. K. Jain, and S. Q. Shen, Phys. Rev. Lett. {\bf102}, 136806 (2009).
\bibitem{Groth2009} C. W. Groth, M. Wimmer, A. R. Akhmerov, J. Tworzydlo, and C. W. J. Beenakker, Phys. Rev. Lett. {\bf103}, 196805 (2009).

\bibitem{Kurz2001} Ph. Kurz, G. Bihlmayer, K. Hirai, and S. Blugel, Phys. Rev. Lett. {\bf86}, 1106 (2001).
\bibitem{Kato2010} Y. Kato, I. Martin, and C. D. Batista, Phys. Rev. Lett. {\bf105}, 266405 (2010).
\bibitem{Yanagihara2002} H. Yanagihara and M. B. Salamon, Phys. Rev. Lett. {\bf89}, 187201 (2002).
\bibitem{Groth2014} C. W. Groth, M. Wimmer, A. R. Akhmerov, and X.Waintal, New J. Phys. {\bf16}, 063065 (2014).
\bibitem{Ndiaye2017a} P. B. Ndiaye, C. A. Akosa, and A. Manchon, Phys. Rev. B {\bf95}, 064426 (2017).
\bibitem{Anderson1958} P. W. Anderson, Phys. Rev. {\bf109}, 1492 (1958).
\bibitem{Ivanova2009} N. B. Ivanova, S. G. Ovchinnikov, M. M. Korshunov, I. M. Eremin, N. V. Kazak, Physics - Uspekhi {\bf52}, 789-810 (2009).
\bibitem{Ning2008} F. L. Ning, S. M. Golin, K. Ahilan, T. Imai, G. J. Shu, and F. C. Chou, Phys. Rev. Lett. {\bf100}, 086405 (2008).
\bibitem{Batista2016} C. D. Batista, S.-Z. Lin, S. Hayami, and Y. Kamiya, Rep. Prog. Phys. {\bf 79}, 084504 (2016).
\bibitem{Shindou2005} R. Shindou and K. I. Imura, Nucl. Phys. B {\bf 720}, 399-435 (2005).
\bibitem{Gradhand2012} M. Gradhand, D. V. Fedorov, F. Pientka, P. Zahn, I. Mertig and B. L. Gy\"{o}rffy. J. Phys: Condens. Matter {\bf 24}, 213202 (2012).
\bibitem{Gardner2010} J. S. Gardner, M. J. P. Gingras, J. E. Greedan, Rev. Mod. Phys. {\bf82}, 53 (2010); Steven T. Bramwell and Michel J. P. Gingras, Science {\bf294}, 1495 (2001).
\bibitem{Ma2015} Ma \etal, Science 350, 538 (2015).
\bibitem{Machida2007} Y. Machida, S. Nakatsuji, Y. Maeno, T. Tayama, T. Sakakibara, and S. Onoda, Phys. Rev. Lett. {\bf98}, 057203 (2007).

\end{thebibliography}
\end{document}